\documentclass[preprint,5p,10pt,oneside,sort&compress,biblatex]{elsarticle}

\usepackage{siunitx}
\usepackage{graphicx}  
\usepackage{booktabs,caption}
\usepackage[flushleft]{threeparttable}
\usepackage{textcomp}
\usepackage[colorlinks=true]{hyperref}

\journal{Journal of Alloy and Compounds}
\begin{document}

\begin{frontmatter}

\title{Stabilization of CeGe$_3$ with Ti and O featuring tetravalent Ce ions: 
(Ce$_{0.85}$Ti$_{0.15}$)Ge$_3$O$_{0.5}$}

\author[ucdphy]{Hanshang Jin}
\author[ucdchem]{Jackson Badger}
\author[ucdphy]{Peter Klavins}
\author[ucdphy,shanghai]{Jing-Tai Zhao}

\author[ucdphy]{Valentin Taufour\corref{mycorrespondingauthor}}
\cortext[mycorrespondingauthor]{Corresponding author}

\address[ucdphy]{Department of Physics, University of California Davis, Davis, California 95616, U.S.A.}
\address[ucdchem]{Department of Chemistry, University of California Davis, Davis, California 95616, U.S.A.}
\address[shanghai]{School of Materials Science and Engineering, Guilin University of Electronic Technology, Guilin 541004, China}

\begin{abstract}
Sub-oxides with the anti-perovskite structure constitute a large part of the interstitial stabilized compounds with novel chemical and physical properties, especially in the systems containing rare earth elements and/or early transition metal elements. A new sub-oxide compound, (Ce$_{0.85}$Ti$_{0.15}$)Ge$_3$O$_{0.5}$, is synthesized by flux growth. The compound crystallizes in a cubic structure which can be considered as an ordered vacancy variant of anti-perovskite-structure or an ordered interstitial variant of the  Cu$_3$Au type, with a space group of Fm-3m (225). The structure is refined from X-ray single crystal methods. The compound may be considered as a compound stabilized by chemical pressure generated by the substitution and the interstitials atoms. The stability of the compounds in RTr$_3$ systems (R = rare earth elements except Pm and Sc, Tr = tetrel elements such as Si, Ge, Sn, Pb) seem to be highly correlated to the ratio of the rare earth elements' radii to the main group elements' atomic radii. The magnetic property of the compound confirms the 4$^+$ valence of the Ce atoms. The resistivity measurement of the compound shows that it is metallic.
\end{abstract}
\begin{keyword}
\texttt{(Ce$_{0.85}$Ti$_{0.15}$)Ge$_3$O$_{0.5}$}\sep CeGe$_3$\sep sub-oxide \sep magnetic property
\end{keyword}

\end{frontmatter}

\section{Introduction}

It is well known that Ce compounds often show novel magnetic properties due to the intermediate valence or heavy fermion behavior of cerium. CeTr$_3$ compounds exist with Tr = Sn~\cite{Sereni:1980fa,Liu:1980fj}, Pb~\cite{Vettier:1986jp,Murani:1983ki,Lin:1985kk,Strange:1986kz,Welp:1987ek} with ambient pressure syntheses. They often show valence fluctuation heavy Fermion behaviour, Kondo effect, etc. No compounds are reported for Tr = Ge and Si at ambient pressure conditions~\cite{Eremenko:ek,osti_4585357,Gokhale:1989cu}. Lanthanoid (Ln) ions with smaller-sized radii such as Gd$^{3+}$, Ho$^{3+}$, Y$^{3+}$, Er$^{3+}$ form in the orthorhombic DyGe$_3$ structure at ambient condition~\cite{Morozkin:1999gj, Physics:wt, Belyavina:1999bm, SchobingerPapamantellos:vz, SchobingerPapamantellos:1992fj}.

High pressure (hp) can induce phase transitions and may lead to new compounds with new structures, especially for elements with valence fluctuations, such as Ce. A useful compilation of the binary high pressure compounds can be found in Ref.~\cite{Demchyna:2006kc}. High pressure synthesis of CeGe$_3$~\cite{Fukuoka:2004ft} was reported under a pressure of \SI{5}{\giga\pascal} at \SI{1600}{\celsius}. It crystallized in a cubic Cu$_3$Au-type structure with a = 4.354(4)\,\si{\angstrom}, isotypic with CeSn$_3$ and CePb$_3$. The specific atomic volume (unit cell volume / total number of atoms in unit cell) is 20.635\,\si{\angstrom}$^3$, which is similar to the newly discovered cubic Cu$_3$Au-type SmGe$_3$ that is also synthesized under high pressure with a specific atomic volume of 20.38\,\si{\angstrom}$^3$~\cite{Hubner:2020jl}.
The high pressure and high temperature synthesis (3 $\sim$ \SI{12}{\giga\pascal}, 500 $\sim$ \SI{1200}{\celsius}) of LaGe$_3$ lead to the formation of a compound with a BaPb$_3$-type instead of Cu$_3$Au-type structure~\cite{Fukuoka:2011ba} with specific atomic volume of 21.78\,\si{\angstrom}$^3$, larger than hp-CeGe$_3$. PrGe$_3$ and NdGe$_3$ form a new type of structure~\cite{Fukuoka:2010ig} with specific atomic volume 20.89\,\si{\angstrom}$^3$ and 21.36\,\si{\angstrom}$^3$ respectively, also larger than hp-CeGe$_3$. The superconducting compounds of LaSn$_3$ and LaPb$_3$ were found to crystallize in the Cu$_3$Au-type structure, similar to CeSn$_3$ and CePb$_3$~\cite{Ram:2013jp,Gambino:1968eh,Harris:1965kk, Strange:1986kza}.

The existence of the hp-CeGe$_3$ phase indicates the possibility to stabilize the structure by chemical pressure, namely by substitution of smaller atoms in the structure. The phase diagram of the ternary cross section at 1170\,K reported by A.V. Morozkin~\cite{Morozkin:2004dn} did not mention any possible (Ce,Ti)Ge$_3$ existence in the region.

The interstitial atoms, such as B, C, N and O, often stabilize compounds with cavities such as in the anti-perovskite structure. Sub-oxides with the anti-perovskite structure constitute a large part of the interstitial stabilized compounds with novel physical properties~\cite{Zhao:1995dg}, especially in the systems containing rare earth elements and/or early transition metal elements, which often form impurity-stabilized phases~\cite{Ferro:2010ub}.

The existence of the title compound may be resulted from both chemical pressure and oxygen stabilization. Here we report the synthesis, structure and physical properties characterizations of (Ce$_{0.85}$Ti$_{0.15}$)Ge$_3$O$_{0.5}$. We find that the compound has the cubic structure similar to hp -CeGe$_3$, with the 4$^+$ valence of the Ce atoms confirmed by magnetization measurements, and with metallic behavior. We suspect that the stability of the compounds in the RTr$_3$ family is closely linked to the ratio of the rare earth elements' radii to the main group elements' radii (r$_{R}$/r$_{Tr}$).

%%%%%%%%%%%%%%%%%%%%%%%%%%%%%%%%%%%%%%%%%%%%%%%%%%%%%%%%%
%%%%%%%%%%%%%%%%%%%%%%%%%%%%%%%%%%%%%%%%%%%%%%%%%%%%%%%%%

\section{Experimental Section}

\begin{figure}[!htp]
\center
\includegraphics[width=0.9\linewidth]{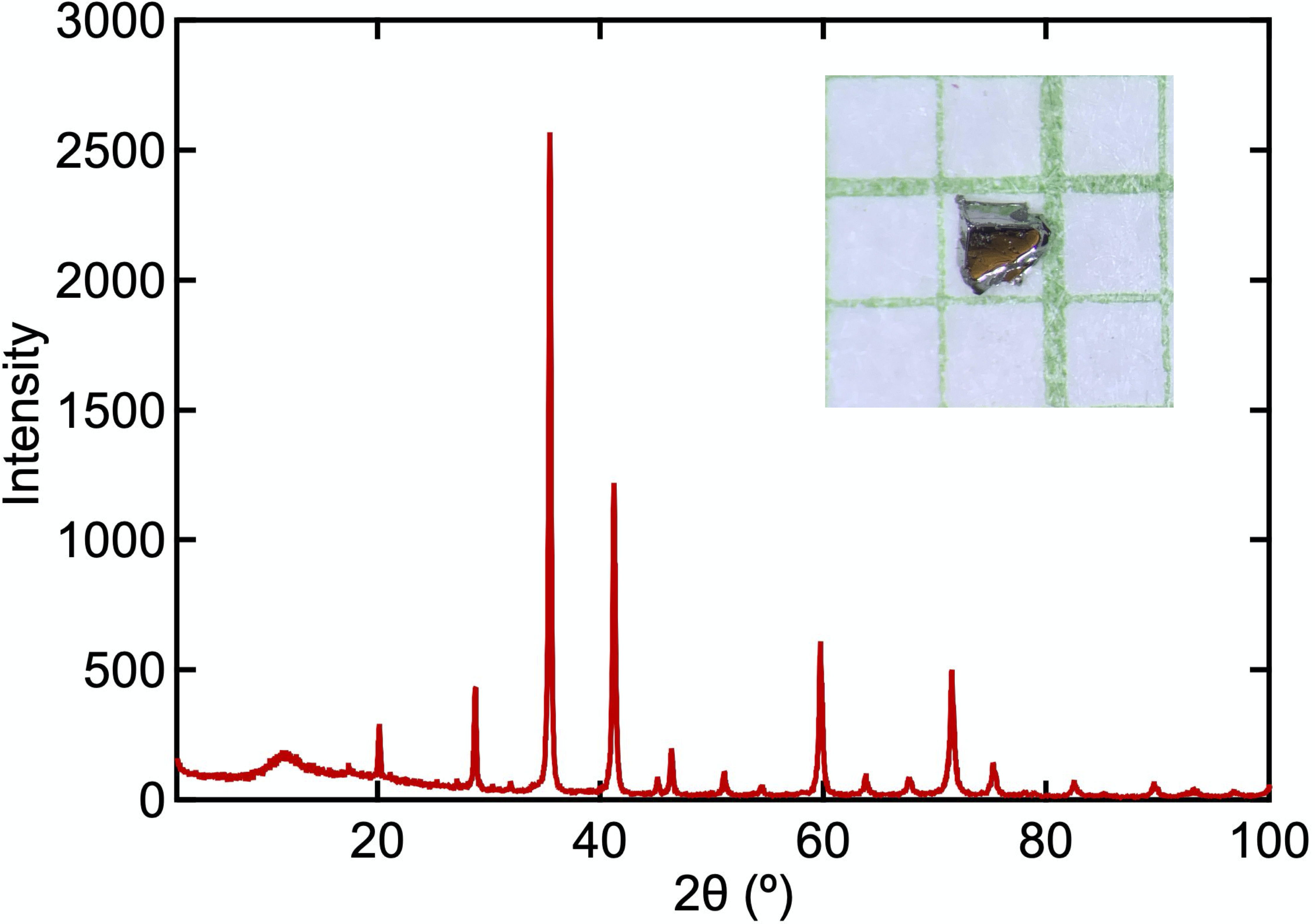}
\caption{X-ray powder diffraction pattern of the sample with the inset photo of a crystal, (Ce$_{0.85}$Ti$_{0.15}$)Ge$_3$O$_{0.5}$.}
\label{fig:crystal_xrd}
\end{figure}

In this work, the single crystal samples were grown from self-flux solution growth. The starting materials [Ce pieces (Ames Lab), Ti granules (4N), Ge lumps (6N)] with initial stoichiometry Ce$_4$Ti$_1$Ge$_{19}$, were introduced in a 2\,mL Canfield Crucible Set~\cite{Canfield:2016cn}, sealed in a fused silica ampoule with 0.2\,atm. of argon, heated to \SI{1200}{\celsius} in 5 hours, held at \SI{1200}{\celsius} for 10 hours, cooled to \SI{1050}{\celsius} in 2.5 hours, held at \SI{1050}{\celsius} for 1 hour, slow cooled to \SI{860}{\celsius} in 146 hours. The ampoule was then taken out of the furnace and quickly spun to separate crystals from the melt. This synthesis procedure was designed for the growth of CeTiGe$_3$ single crystals which have interesting properties related to magnetic quantum criticality~\cite{Kaluarachchi:2018if}. Most of the crystals were single crystals of CeTiGe$_3$ which have a characteristic thick hexagonal plates morphology.
In addition, we found small amounts of crystals with a cubic morphology: (Ce$_{0.85}$Ti$_{0.15}$)Ge$_3$O$_{0.5}$. It was not our intention to target this new material. The oxygen necessary to form this phase most likely originated from a slight oxidation of the Ce. Although an oxide contamination often forms an insoluble layer floating on the surface of intermetallic melts, in the case of this Ce-Ti-Ge melt, oxygen is clearly soluble to some degree. The crystals have normally regular shapes with shining surfaces, as shown in the inset of Fig.~\ref{fig:crystal_xrd}, with the powder X-ray diffraction pattern of the crystal.

In the Ref.~\cite{Kaluarachchi:2018if}, the temperature profile is first heated to \SI{1200}{\celsius} and then slowly cooled to \SI{900}{\celsius} over 120h with a cooling rate is \SI{2.5}{\celsius}/h, and there is no report of this cubic impurity phase. In a different paper that also synthesizes CeTiGe$_3$ with flux method~\cite{Inamdar:2014io}, their temperature profile is first heated to \SI{1050}{\celsius} and then slowly cooled to \SI{850}{\celsius} with an unspecified cooling rate, and there is also no report of the cubic impurity phase. Compared with the above two temperatures, we were able to spend a lot more time in the growth region(from \SI{1050}{\celsius} to \SI{860}{\celsius}) with a cooling rate of \SI{1.3}{\celsius}/h. Given that the size of the cubic phase crystal is relatively large, it is unlikely that it forms during the quick cooling from \SI{1200}{\celsius} to \SI{1050}{\celsius}. And probably because of the slower cooling rate below \SI{1050}{\celsius}, we were able to observe this phase where all three synthesis have the same initial composition.

In the later experiments, we were able to avoid this cubic phase by changing the initial stoichiometry ratio to Ce:Ti:Ge = 15:6:79. Based on these results, we suspect that Ref.~\cite{Kaluarachchi:2018if} and ~\cite{Inamdar:2014io} should have formed the cubic phase, but the crystals were not mentioned, probably because the faster cooling rate made these crystals less likely to be well formed.

The phase identifications of the as-obtained samples were carried out by powder X-ray diffraction (PXRD) on a Rigaku Miniflex 600 diffractometer with Cu K$\alpha$ ($\lambda$ = 1.54178\,\si{\angstrom}) radiation at room temperature. The powder pattern contributed by the main phase can be indexed by a cubic cell about twice as large as the anti-perovskite cell. This obvious discrepancy from the Cu$_3$Au structure enhanced our interest to investigate its crystal structure. 

Single-crystal X-ray diffraction data were collected on several samples at 100\,K using a Bruker Photon100 CMOS X-ray diffractometer (Bruker AXS). A sealed-tube Mo X-ray source was employed.

Scanning Electron Microscope (SEM) images and elemental composition were obtained for selected samples on a Hitachi S-4100T Hitachi HTA America with an Oxford INCA Energy Dispersive Spectrometer (EDS). Point scans and elemental mappings were collected to determine the atomic composition across several crystals, with 3-5 spectra collected for each crystal. The atomic percentages were normalized to the germanium atomic percentage to determine the average formula unit across all the spectra.

Magnetic properties were characterized by a SQUID magnetometer (MPMS, Quantum Design) with 7\,T magnetic field, in the temperature range of 2\,K to 300\,K. Resistivity measurements were carried out by a Quantum Design Physical Property Measurement System from 1.8\,K to 300\,K. The ac resistivity ($f=17$\,Hz) was measured by the standard four-probe method using Pt wires (0.002 inch diameter) with silver-filled epoxy, applying a 1\,mA current.

%%%%%%%%%%%%%%%%%%%%%%%%%%%%%%%%%%%%%%%%%%%%%%%%%%%%%%%%%
%%%%%%%%%%%%%%%%%%%%%%%%%%%%%%%%%%%%%%%%%%%%%%%%%%%%%%%%%

\section{Results and Discussion}
\subsection{Crystal Structure} 
\begin{figure}[!htp]
\center
\includegraphics[width=1\linewidth]{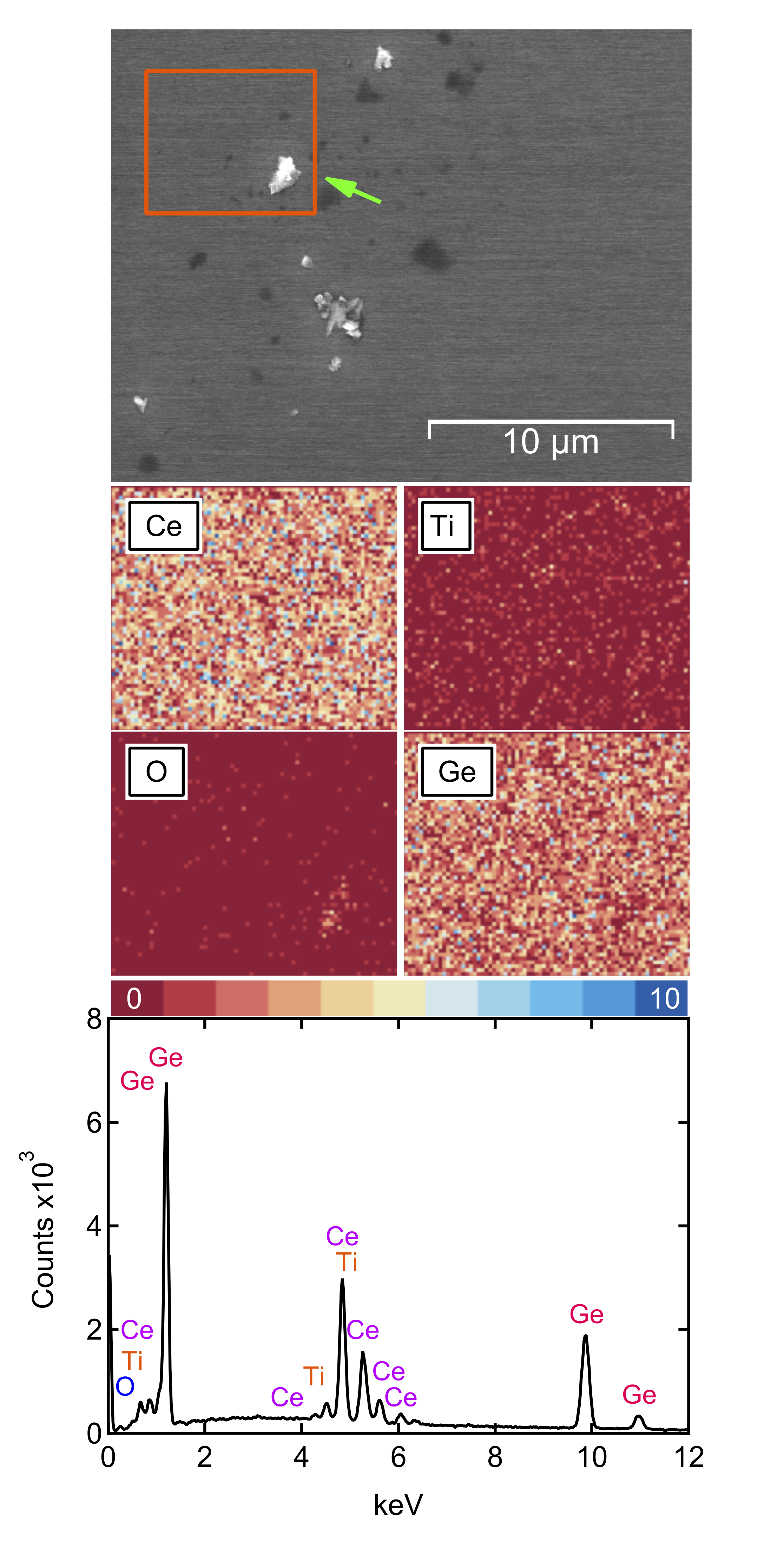}
\caption{SEM image of a bulk crystal with an elemental mapping performed over a selected area, orange rectangle, to determine the uniformity of Ce, Ti, O, and Ge. The arrow shows the pebble (white spot) on the surface of the crystal with a slight excess of oxygen. From the selected area, an EDS spectrum shows the successful identification of each element.}
\label{fig:crystal_edx}
\end{figure}

A SEM image showing the surface of a crystal is shown in Fig. ~\ref{fig:crystal_edx}. As can be seen in this image, the surface of the bulk crystal shows a near uniform topology with a few pebbles on the surface, indicated by the white spots. To determine the composition of the bulk crystal and these pebbles, an elemental mapping was performed over an area, as shown by the orange rectangle. This area reveals a complete homogenous mixture of the elements with a slight excess of oxygen in the pebble, as highlighted by the green arrow and cluster within the oxygen mapping in the bottom right corner. Thus, indicating a uniform substitution of Ti and doping of O across the bulk crystal. From this elemental mapping an exemplary EDS spectrum shows the successful identification of cerium, titanium, germanium, and oxygen. After normalization, the average formula unit across all the spectra is calculated to be (Ce$_{0.89}$Ti$_{0.11}$)Ge$_{3}$O$_{x}$, which is quite similar to the refined value from the single crystal X-ray diffraction analysis described below. Given the inherent difficulties in determining a quantitative oxygen composition from this technique, we have decided to place an $x$ in the formula unit to show the uncertainty of this composition.

The crystal structure was firstly solved and refined using WinCSD package in space group Fm-3m~\cite{Akselrud:tm}. Preliminary refinement shows abnormal behavior on Ge displacement factor, which lead to two considerations: an ordering of the Ge atoms in a lower symmetry space group or a displacement in a disordered way to accommodate the oxygen atoms comfortably. The first possibility was evaluated by refining in all the possible lower symmetry space groups which lead to similar results as obtained from high symmetry refinements. Although we cannot rule out the possibility of a lower symmetry structure, the structure refinements give no clear distinction between the choices of different space groups. Additionally, the super structural nature makes the percentage of observed points much less than a normal structure which lead the refinement at a lower symmetry even more unstable. The attempted refinement of the oxygen occupancy did not lead to better results therefore the occupancy of O position was fixed at 0.5 in the final refinement.

\begin{figure}[!htp]
\centering
\includegraphics[width=0.7\linewidth]{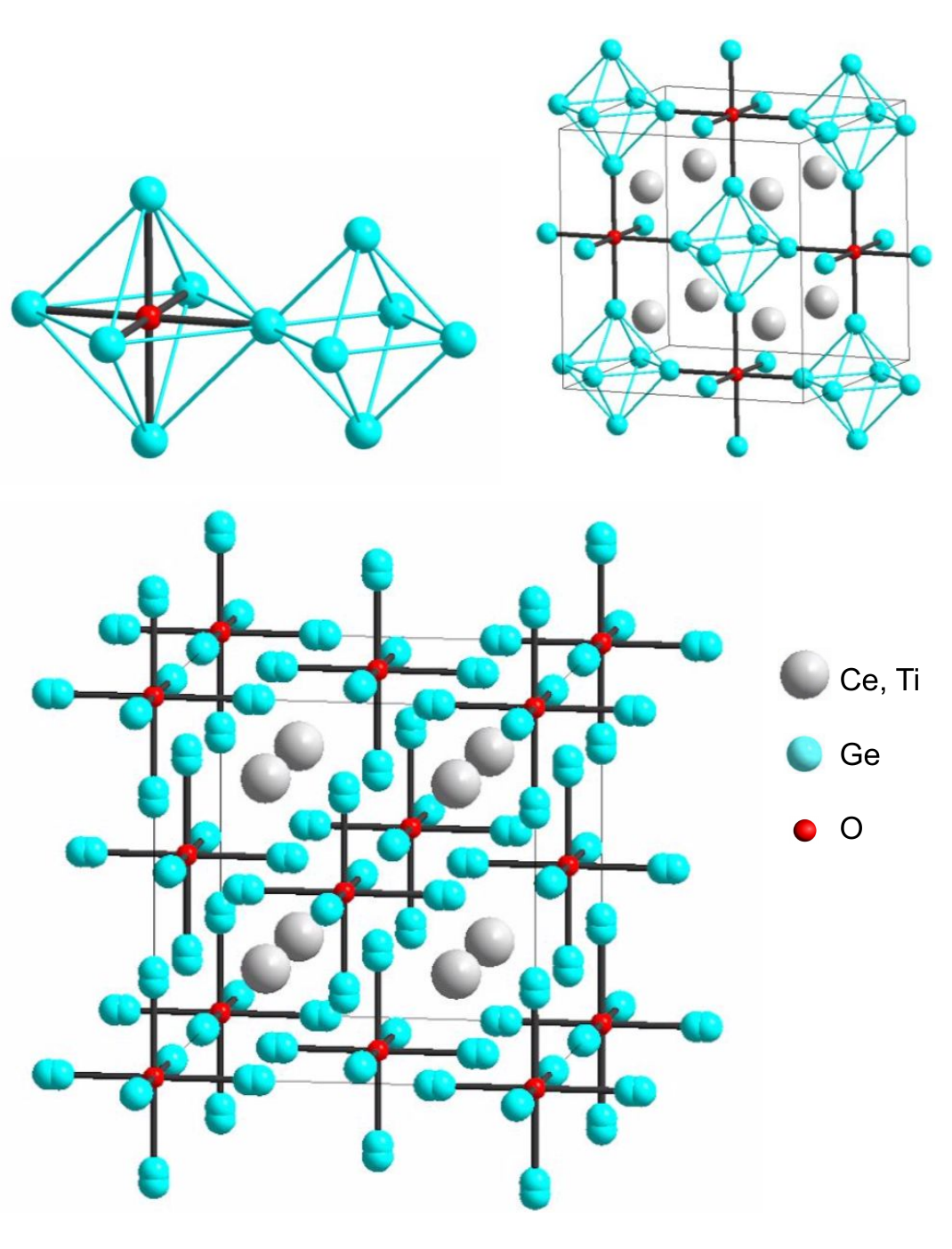}
\caption{Blue, red and grey atoms are Ge, O and the mixture of Ce and Ti atoms, respectively. Ge atoms forms two kinds of octahedron with the larger one centered by O atom and smaller ones left empty (upper left), the two kinds of octahedra arranged in the way by sharing corners alternatively (upper right). The mixture of Ce and Ti atoms occupy the larger cavities in the Ge polyhedral network, and the splitting Ge atoms are also shown in the lower figure.}
\label{fig:cryst_str}
\end{figure}

The final refinements were then carried out in the highest symmetry space group of Fm-3m by splitting the Ge position into two. This lead to a better result which is structurally more reasonable. The oxygen atoms occupy the Ge octahedral interstitial site in an ordered way, while the Ge atoms modify their positions making two kinds of octahedron considerably different from each other, one with Ge-Ge distance of $\sim$2.74\,\si{\angstrom} and the other of $\sim$3.44\,\si{\angstrom}. The larger cavity accommodates the O atom with Ge-O distance of $\sim$2.17\,\si{\angstrom}, which is about 10\% larger than the sum of the Ge and O covalent radii. Similar tendency with distances were observed in some anti-perovskite compounds with C, N and O interstitials residing in rare earth cavities~\cite{Zhao:1995dg}. 
The other much smaller cavity was left empty. The atomic arrangement in the unit cell and the Ge$_6$O polyhedral arrangement are shown in Fig.~\ref{fig:cryst_str}, and the crystal structure refinement results, atomic positional and displacement parameters and atomic distances are listed in Table.~\ref{cs_refine}, Table.~\ref{position} and Table.~\ref{atomic_dist}. 

\begin{table}[!htbp]
\centering
\caption{Crystal structure refinement results of (Ce$_{0.85}$Ti$_{0.15}$)Ge$_3$O$_{0.5}$}
\label{cs_refine}
\resizebox{\columnwidth}{!}{
\begin{tabular}{|ll|}
\hline
Refined items                                               			& Values                        \\ \hline
\multicolumn{1}{|l|}{Chemical formula}                      		& Ce$_{0.85}$Ti$_{0.15}$Ge$_3$O$_{0.5}$            \\
\multicolumn{1}{|l|}{Chemical formula weight}               	& 352.26                      \\
\multicolumn{1}{|l|}{Crystal system}                        		& cubic                         \\
\multicolumn{1}{|l|}{Space group}                           		& Fm-3m                         \\
\multicolumn{1}{|l|}{Cell length $a$ (\si{\angstrom})}               	& 8.7364(8)                    \\
\multicolumn{1}{|l|}{Cell volume (\si{\angstrom}$^3$)}         	& 666.8(2)                     \\
\multicolumn{1}{|l|}{Z}                                     			& 8           \\
\multicolumn{1}{|l|}{D$_x$ (g/cm$^3$)}                            		& 7.0170                        \\
\multicolumn{1}{|l|}{F(000)}                                			& 153                       \\
\multicolumn{1}{|l|}{Absorption coefficient (cm$^{-1}$)}     	& 399.3                         \\
\multicolumn{1}{|l|}{Measurement temp. ($K$)}           	& 295.0                         \\
\multicolumn{1}{|l|}{Monochromator}                         		& Graphite Monochromator        \\
\multicolumn{1}{|l|}{Detector}                              			& CCD                           \\
\multicolumn{1}{|l|}{Theta range ($^\circ$)}                       		& 27.6                          \\
\multicolumn{1}{|l|}{Radiation wavelength (\si{\angstrom})}            		& 0.710730                      \\
\multicolumn{1}{|l|}{Radiation type}                        		& Mo K$\alpha$                          \\
\multicolumn{1}{|l|}{Reflections number}                         		& 588                           \\
\multicolumn{1}{|l|}{R$_{int}$}                                  		& 0.0463                        \\
\multicolumn{1}{|l|}{av$\_\sigma$(I)/netI}                          	& 0.0270                        \\
\multicolumn{1}{|l|}{Number (total)}                         		& 58                            \\
\multicolumn{1}{|l|}{Number (gt)}                           		& 58                            \\
\multicolumn{1}{|l|}{Threshold expression}                 		& F$_o$ $>$4$\sigma$(F$_o$)                   \\
\multicolumn{1}{|l|}{Program package}                       		& WinCSD                        \\
\multicolumn{1}{|l|}{Weighting details}                     		& 0.1285Log(F$_o$)$^4$                \\
\multicolumn{1}{|l|}{Extinction: Becker-Coppens}             	& \begin{tabular}{lll}Ion method \\ Gauss (secondary)\end{tabular} \\
\multicolumn{1}{|l|}{RMS mosaic spread (sec)}               	& 18.66                         \\
\multicolumn{1}{|l|}{Extinction coef.}                      		& 0.0404(3)                     \\
\multicolumn{1}{|l|}{Number reflns}                         		& 59                            \\
\multicolumn{1}{|l|}{R factor (gt)}                         		& 0.0421                        \\
\multicolumn{1}{|l|}{wR factor}                             		& 0.0425                        \\
\multicolumn{1}{|l|}{Goodness of fit}                       		& 1.10                          \\
\multicolumn{1}{|l|}{Residual e-density peak/hole} & 0.79/-0.65 (e$\si{\angstrom}^3$)                  \\ \hline
\end{tabular}}
\end{table}

%%%%%%%%%%%%%%%%%%%%%%%%%%%%%%%%%%%%%%%%%%%%%%%%%%%%%%%%%%%%%%%%%%%%%%%%%%%%%
\begin{table}[!htbp]
\centering
\resizebox{\columnwidth}{!}{
\begin{threeparttable}
	\centering
\caption{Refined atomic positional and displacement parameters of (Ce$_{0.85}$Ti$_{0.15}$)Ge$_3$O$_{0.5}$}
	\label{position}
	\begin{tabular}{|l|l|l|l|l|l|l|}
	\hline
	Name	 & Type	 & Wyckoff 	& x	 & y	 & z	 & B$_{av}$	 \\ \hline
	Ce1* & Ce,Ti & 8(c) & 1/4 & 1/4 & 1/4 & 0.43(7) \\ \hline
	Ge1 & Ge & 24(e) & 0.278(8) & 0 & 0 & 0.8(8) \\ \hline
	Ge2 & Ge & 24(e) & 0.248(8) & 0 & 0 & 1.2(6) \\ \hline
	O1 & O & 4(a) & 0 & 0 & 0 & 1.5(8) \\ \hline
	\end{tabular}
\begin{tablenotes}
\item
Ce1* = Ce$_{0.85(4)}$Ti$_{0.15(4)}$
\end{tablenotes}

\bigskip

\centering
\begin{tablenotes}
\item
Anisotropic displacement parameters:
\end{tablenotes}
\begin{tabular}{|l|l|l|l|l|l|l|}
\hline
Atom & $\beta_{11}$ & $\beta_{22}$ & $\beta_{33}$ & $\beta_{12}$ & $\beta_{13}$ & $\beta_{23}$ \\ \hline
Ce1 & 0.43(12) & 0.43(12) & 0.43(12) & 0 & 0 & 0 \\ \hline
Ge1 & 2.0(21) & 0.2(5) & 0.2(5) & 0 & 0 & 0 \\ \hline
Ge2 & 1.0(14) & 1.3(7) & 1.3(7) & 0 & 0 & 0 \\ \hline
\end{tabular}
\end{threeparttable}}

\end{table}

% \usepackage{graphicx}
\begin{table}[!htbp]
\fontsize{11}{11}
\centering
\caption{Refined atomic distances($\si{\angstrom}$) of (Ce$_{0.85}$Ti$_{0.15}$)Ge$_3$O$_{0.5}$}
\label{atomic_dist}
\begin{tabular}{|l|l|}
\hline
Ge1 - O1  & 2.43(7)   \\
Ge2 - O1  & 2.17(7)   \\
Ge1 - Ge1 & 2.74(7)   \\
Ge1 - Ge2 & 2.93(7)   \\
Ge2 - Ge2 & 3.07(7)   \\
Ge2 - Ge2 & 3.11(7)   \\
Ge1 - Ge2 & 3.26(7)   \\
Ge1 - Ge1 & 3.44(7)   \\
Ce1 - Ge2 & 3.0888(3) \\
Ce1 - Ge1 & 3.099(6)  \\
Ce1 - Ce1 & 4.3682(2) \\ \hline
\end{tabular}
\end{table}

The specific atomic volume per atom of the title compound is 18.52\,\si{\angstrom}$^3$ (20.83\,\si{\angstrom}$^3$ if ignore oxygen), smaller than 21.78\,\si{\angstrom}$^3$, 20.89\,\si{\angstrom}$^3$ and 21.36\,\si{\angstrom}$^3$ for the high pressure phase of LaGe$_3$, PrGe$_3$ and NdGe$_3$ prepared under similar pressure conditions ( 3 $\sim$ \SI{12}{\giga\pascal}, 500 $\sim$ \SI{1200}{\celsius}), respectively~\cite{Fukuoka:2011ba,Fukuoka:2010ig}. The value for high pressure CeGe$_3$ is 20.64\,\si{\angstrom}$^3$, and for high pressure SmGe$_3$ is 20.38\,\si{\angstrom}$^3$. Both are similar to that of the title compound (ignoring oxygen). This indicates that the substitution of Ti to Ce and extra bonding from oxygen interstitial to germanium atoms resulted to similar shrinkage of the cell comparing to the high pressure, which can be considered as a chemical pressure effect. In consideration of the value of 20.64\,\si{\angstrom}$^3$ for hp-CeGe$_3$, the result might already indicate that the behavior of the Ce is different from La, Pr or Nd, due to the valence change from Ce$^{3+}$ to Ce$^{4+}$ caused by high pressure or chemical pressure. This result is also consistent with our magnetic measurements of the Ti substituted sample, discussed in the later paragraph.

\begin{figure*}[!htp]
\center
\includegraphics[width=14cm]{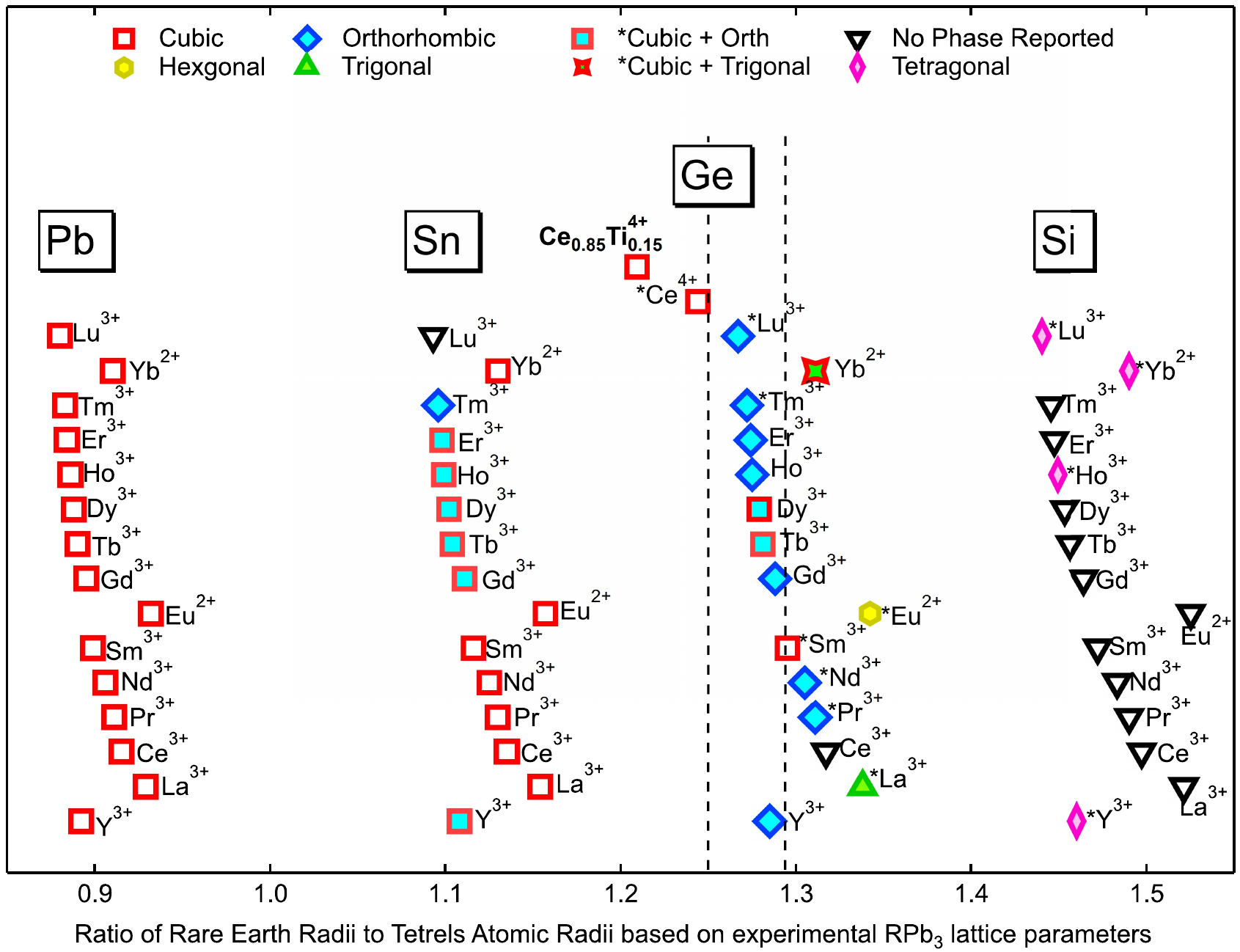}
\caption{Correlation of RTr$_3$ formation and structure types with the ratio of the R's radii to Tr atomic radii. Compounds with star(*) labels are reported in high pressure synthesis.}
\label{fig:correlation}
\end{figure*}

The ratio of the R's radii to Tr atomic radii in the RTr$_3$ phases are presented in Figure~\ref{fig:correlation}. The radii of R was obtained empirically by using the RPb$_3$ series as a reference. The sum of r$_R$ and r$_{Tr}$ was calculated based on their respective cubic lattice parameter, and the empirical radii of R was estimated by subtracting the atomic radii of Tr~\cite{Slater:2004fm}. The radius of Ce$^{4+}$ was estimated based on the lattice parameter of $\alpha$-Ce~\cite{Zachariasen:ua} and a correction factor of 0.902 calculated from the radius of Ce$^{3+}$ in $\gamma$-Ce~\cite{Zachariasen:ua} and the estimated radius in CePb$_3$. The radius of Ti$^{4+}$ was also obtained from its atomic radius~\cite{Slater:2004fm} using the same correction factor. And the radius of Ce$_{0.85}$Ti$_{0.15}^{4+}$ is simply obtained by 85\% of the Ce$^{4+}$ radius with 15\% of Ti$^{4+}$ radius.

As can be seen in Fig.~\ref{fig:correlation}, for the small ratios, the compounds adopt the cubic Cu$_3$Au-type structure, whereas for the large ratios, fewer RTr$_3$ phases are reported.

The RSn$_3$ and RPb$_3$ compounds, having smaller ratios, normally adopt the cubic Cu$_3$Au-type structure. We note that there are a cluster of RSn$_3$ (R = Gd -- Er, Y) compounds, TbGe$_3$ and DyGe$_3$ that can be formed in the cubic structure under pressure synthesis, and in the orthorhombic structure at normal pressure synthesis~\cite{Skolozdra:1986ub, Meier:2010dk, Eremenko:vb, Miller:ul, Palenzona:1993dz, Tsvyashchenko:2013dc}.

The RSi$_3$ compounds, having the large ratios, have not been reported except a small cluster of RSi$_3$ compounds, R = (Ho, Y, Yb and Lu), which form in a tetragonal structure from high pressure synthesis~\cite{Schwarz:2012ia, Wosylus:2011hh, Meier:2011gd}.

The RGe$_3$ family is in between the RSn$_3$ which is mostly cubic and RSi$_3$ which is mostly unreported. For larger rare earth elements, RGe$_3$ phases mostly form in an orthorhombic structure, and some of them are reported only from high pressure synthesis. LaGe$_3$~\cite{Fukuoka:2011ba}, PrGe$_{3.36}$ and NdGe$_{3.25}$~\cite{Fukuoka:2010ig} do not adopt Cu$_3$Au type but hp-CeGe$_3$ does~\cite{Fukuoka:2004ft}. That is probably because of the smaller compressibility of the La, Pr and Nd compared with Ce atoms, due to the lack of valence change. The other three compounds form different structure types and keep the specific atomic volume larger than those values of hp-CeGe$_3$ and (Ce,Ti)Ge$_3$O$_{0.5}$ which resulted from the Ce valence change from 3$^+$ to 4$^+$. In this case, the radii ratio can be estimated as 1.244 for hp-CeGe$_3$ and 1.209 for the title compound. This means the ratio for hp-CeGe$_3$ phase and substituted phase can be considered to have been tuned to be in the range of the Cu$_3$Au type formation, e.g. $<$ 1.25.

We note that for Eu and Yb compounds, the previous reports indicate that EuPb$_3$~\cite{Dai:2013jg}, YbPb$_3$~\cite{Baranovskiy:2004jr}, EuSn$_3$~\cite{Akinobu:2014dr} and YbSn$_3$~\cite{Sakamoto:1997ig} have valence of 2+, and EuGe$_3$~\cite{Castillo:2015gc} and YbGe$_3$~\cite{PeterSebastian:2010cd} have an average valence above 2+. While there is no report of the Yb valence in YbSi$_3$, we expect that it should also be above 2+ given that it is synthesized only under high pressure. 

It is interesting to notice that the existence of the phases and the structure types are highly correlated with the ratio. It seems that the ratio of the rare earth elements radii to the carbon group elements radii is one of the key factors of the formation in different structure types.

We observe from Figure~\ref{fig:correlation} that systems with the ratio \,$>$\,1.294 more likely require the assistance of high pressure synthesis. We also observe that systems with a smaller ratio, especially if the ratio is smaller than 1.25, are more likely to form a Cu$_3$Au type of structure, and some of them require the assistance of high pressure synthesis to adopt the cubic phase. Based on the general trend shown in the figure, we may speculate that TmSn$_3$ and GdGe$_3$ also might have a cubic phase under high pressure synthesis.

\subsection{Magnetic and Resistivity Measurements}
\begin{figure}[!htp]
\centering
\includegraphics[width=0.8\linewidth]{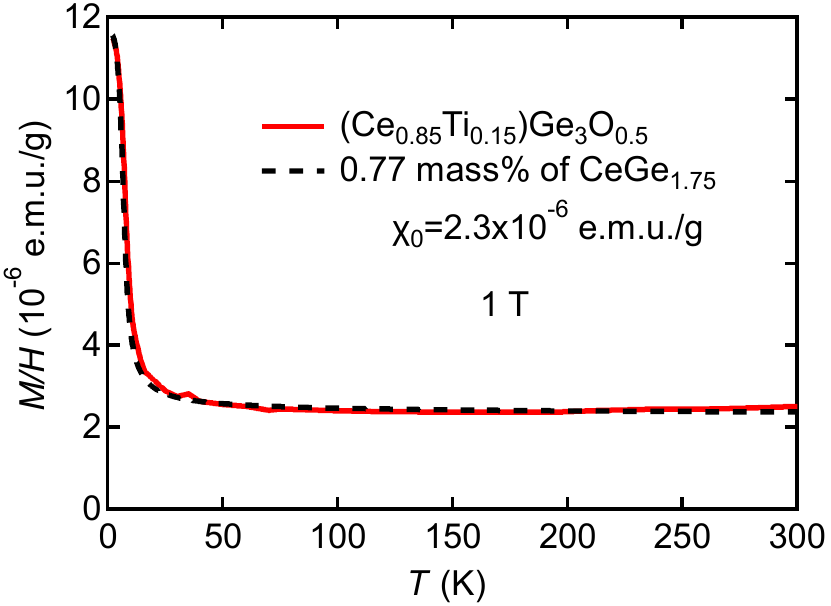}
\includegraphics[width=0.8\linewidth]{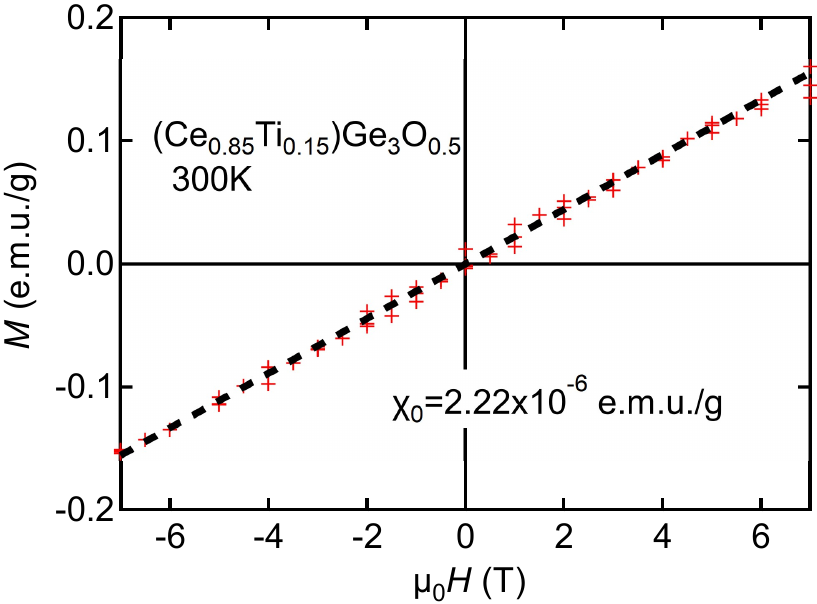}
\caption{(a) Temperature dependence of the magnetic susceptibility and (b) field dependence of the magnetization of (Ce$_{0.85}$Ti$_{0.15}$)Ge$_3$O$_{0.5}$. Red lines are the measurements and black dashed lines are (a) 0.77 mass\% of CeGe$_{1.75}$ and (b) linear fit of the magnetization measurement.}
\label{fig:magnetic}
\end{figure}

Figure~\ref{fig:magnetic} presents the temperature dependence of the magnetization of the title compound. The magnetization is relatively temperature independent above $\sim20$\,K indicating a Pauli paramagnetic behavior. We observe a larger magnetization at lower temperature which we attribute to trace amounts of CeGe$_{1.75}$. This compound orders ferromagnetically below $7$\,K~\cite{Budko:2014gf}. All our single crystals show traces of CeGe$_{1.75}$. To subtract its contribution, we prepared a pure sample of CeGe$_{1.75}$ and measured its magnetization under a field of $1$\,T. We then fit the magnetization of (Ce$_{0.85}$Ti$_{0.15}$)Ge$_3$O$_{0.5}$ by assuming a constant magnetization in addition to a small percentage of CeGe$_{1.75}$. The dash line in Fig.~\ref{fig:magnetic} shows a good fit with only $0.77$ mass\% of CeGe$_{1.75}$ impurity. We find the magnetic susceptibility of (Ce$_{0.85}$Ti$_{0.15}$)Ge$_3$O$_{0.5}$ to be $\chi_0 = 2.3\times10^{-6}$\,e.m.u./g. This is also consistent with the slope of the magnetization versus field at $300$\,K shown in Figure~\ref{fig:magnetic}(b). This Pauli paramagnetic behavior implies that the Ce valence is $4^+$. 

\begin{figure}[!htp]
\centering
\includegraphics[width=0.8\linewidth]{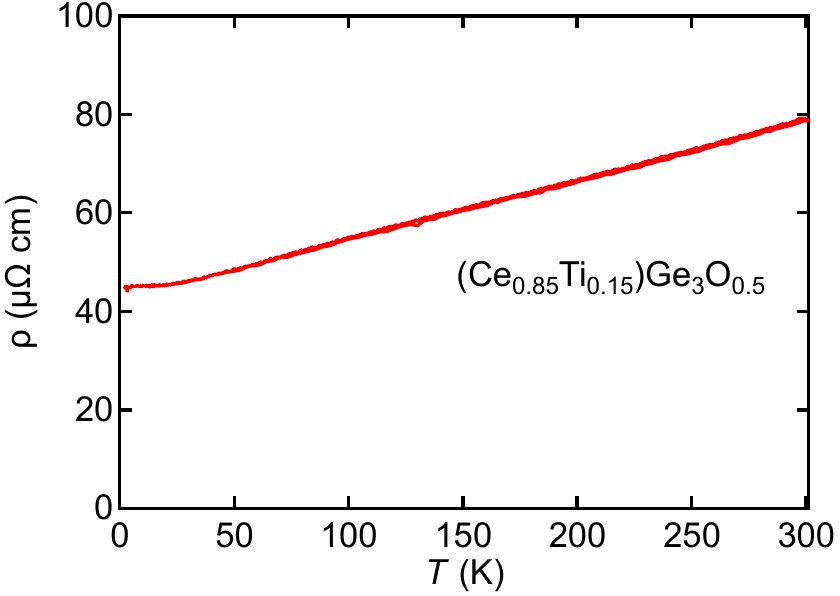}
\caption{Temperature dependence of the electrical resistivity for (Ce$_{0.85}$Ti$_{0.15}$)Ge$_3$O$_{0.5}$ from 2K to 300K.}
\label{resis}
\end{figure}

Figure~\ref{resis} presents the temperature dependence of the electrical resistivity  of (Ce$_{0.85}$Ti$_{0.15}$)Ge$_3$O$_{0.5}$. It shows a metallic behavior, with $79\,\mu\Omega\cdot$cm at $300$\,K and residual resistivity of $45\,\mu\Omega\cdot$cm at $2$\,K. The resistivity of high pressure synthesized CeGe$_3$~\cite{Fukuoka:2004ft} and CeSn$_3$~\cite{Sereni:1980fa} are respectively $130$ and $35\,\mu\Omega\cdot$cm at $300$\,K, and the residual resistivity are respectively $7$ and $0.7\,\mu\Omega\cdot$cm. Compared to the hp-CeGe$_3$, it is expected that the (Ce$_{0.85}$Ti$_{0.15}$)Ge$_3$O$_{0.5}$ has higher resistivity at low temperature due to the impurity scattering from titanium and oxygen atoms. But it is unknown why it has lower resistivity at higher temperature. The Matthiessen's rule may not be valid, given that the compound has a fairly large amount of substitution~\cite{Dube:1938ff,Kaveh:1986ir}.

%%%%%%%%%%%%%%%%%%%%%%%%%%%%%%%

%%%%%%%%%%%%%%%%
\section{Conclusion}
The compound (Ce$_{0.85}$Ti$_{0.15}$)Ge$_3$O$_{0.5}$ is reported. It crystallizes in a variant of the Cu$_3$Au-type structure. The compound may be considered as a stabilization by chemical pressure generated by the Ce-Ti substitution and the O-interstitials atom. The stability of the compounds in the RTr$_3$ systems are highly correlated to the ratio of the atomic radii of rare earth elements to the carbon group elements (r$_{R}$/r$_{Tr}$). In an analogous situation, we may speculate that systems with ratio $>$ 1.294 likely require high pressure synthesis to form compounds. The RTr$_3$ phases with a smaller ratio, especially smaller than 1.25, are more likely to form a Cu$_3$Au type of structure. The magnetic property of the compound confirms the 4$^+$ valence of the Ce atoms. The resistivity measurement shows that it is metallic, similar to other CeTr$_3$ compounds.

%%%%%%%%%%%%%%%%%%%%%%%%%%%%%%%
%%%%%%%%%%%%%%%%%%%%%%%%%%%%%%%%%%%%%%%%%%%%%%%%%%%%%%%%%%%%%%%%%%%%%%%%%%%%%

\section{Acknowledgement}
This work is financially supported by the Physics Department, University of California, Davis. Jing-Tai Zhao is grateful to Professor Juri Grin from the Max-Planck Institute for Chemical Physic of Solids for fruitful discussions on crystal structure refinements.

%%%%%%%%%%%%%%%%%%%%%%%%%%%%%%%%%%%%%%%%%%%%%%

\bibliography{Cubic_paper}

%%%%%%%%%%%%%%%%%%%%%%%%%%%%%%%%%%%%%%%%%%%%%%

% \usepackage{graphicx}

%%%%%%%%%%%%%%%%%%%%%%%%%%%%%%%%%%%%%%%%%%%%

\end{document}